\documentstyle[editedvolume,psfig]{crckapb}

\def\lsim{\lower.5ex\hbox{$\; \buildrel < \over \sim \;$}}
\def\gsim{\lower.5ex\hbox{$\; \buildrel > \over \sim \;$}}

\begin{opening}
\title{LiBeB ENERGETICS AND COSMIC RAY ORIGIN}

\author{Reuven Ramaty}
\institute{NASA/GSFC, Greenbelt, MD 29771, USA}

\author{Richard E. Lingenfelter}
\institute{UCSD, LaJolla, CA 92093, USA}

\end{opening}

\runningtitle{LiBeB AND COSMIC RAY ORIGIN}

\begin{document}

\begin{abstract}

Three different models have been proposed for LiBeB production by
cosmic rays: the CRI model in which the cosmic rays are
accelerated out of an ISM of solar composition scaled with
metallicity; the CRS model in which cosmic rays with composition
similar to that of the current epoch cosmic rays are accelerated
out of fresh supernova ejecta; and the LECR model in which a
distinct low energy component coexists with the postulated cosmic
rays of the CRI model. These models are usually distinguished by
their predictions concerning the evolution of the Be and B
abundances. Here we emphasize the energetics which favor the CRS
model. This model is also favored by observations showing that the
bulk (80 to 90\%) of all supernovae occur in hot, low density
superbubbles, where supernova shocks can accelerate the cosmic
rays from supernova ejecta enriched matter.

\end{abstract}

\section{Introduction and Overview}

In a series of papers (Ramaty et al. 1997; Ramaty, Kozlovsky \&
Lingenfelter 1998a; Lingenfelter, Ramaty \& Kozlovsky 1998; Higdon,
Lingenfelter \& Ramaty 1998; Ramaty, Lingenfelter \& Kozlovsky
1998b) we developed a LiBeB (Li, Be, B) and cosmic-ray origin
paradigm (CRS) in which at all epochs of Galactic evolution the
cosmic rays are accelerated out of fresh supernova ejecta, and all
of the Be and part of the B are produced by interactions of such
cosmic rays with the ambient interstellar medium (ISM). This model
differs from the CRI paradigm which posits that the current epoch
cosmic rays are accelerated out an ambient medium of solar
composition (suggested to be the ISM, Meyer, Drury \& Ellison 1997),
and that at all past epochs the composition of the source particles
of the cosmic rays was that of the average ISM at that epoch. Hybrid
models of LiBeB origin were also suggested (Cass\'e, Lehoucq \&
Vangioni-Flam 1995; Vangioni-Flam et al. 1996; Ramaty, Kozlovsky \&
Lingenfelter 1996). In these models (LECR) a Galaxy-wide separate
low energy cosmic-ray component, also accelerated out of fresh
nucleosynthetic matter, coexists with the CRI cosmic rays and
dominates the Be and B production, particularly in the early Galaxy.

The excess of the observed Be abundances in low metallicity stars
over the CRI predictions was discussed by Pagel (1991), and as
additional Be data accumulated (see Vangioni-Flam et al. 1998), it
became clear that the dependence of log(Be/H) on [Fe/H] is
essentially linear, not quadratic as predicted by this model. On the
other hand, both the CRS and LECR models predict a linear evolution,
consistent with the observations, although Fields \& Olive (1998)
recently suggested that the CRI model should still be considered as
viable, based on their re-analysis of the data including O in low
metallicity stars (Israelian, Garcia Lopez, \& Rebolo 1998).

We showed previously (see references above) that the energy
$W_{\rm SN}$ in cosmic rays per supernova required to produce the
observed Be abundance is a powerful diagnostic of the models. This
can be seen by considering log(Be/Fe) as a function of [Fe/H] in
Figure~1a (Ramaty et al. 1998a; Vangioni-Flam et al. 1998), where
for [Fe/H]$ < -1$ the data are consistent with a constant, ${\rm
log(Be/Fe)}=-5.84\pm0.05$. Since Fe production in this epoch is
dominated by core collapse supernovae (SNII), the constancy of
Be/Fe strongly suggests that Be production is also due to SNIIs,
which is eminently reasonable since supernova shocks are the most
likely accelerators of the cosmic rays (e.g. Axford 1981). The
decrease of Be/Fe for [Fe/H]$\gsim -1$ probably results from the
additional Fe production in Type Ia supernovae (e.g. Matteucci \&
Greggio 1986). The essentially constant Be/Fe, together with
information on the average Fe yield per SNII, allows the
determination of the Be yield per SNII which, coupled with
calculations of LiBeB production by cosmic rays (Ramaty et al.
1997), leads to the energy in cosmic rays per SNII for the various
models. We have shown that for the CRS model $W_{\rm SN} \simeq
10^{50}$erg, a value which is quite consistent with that required
to produce the current epoch cosmic rays, based on direct cosmic
ray measurements and supernova statistics. We have also shown that
the LECR model is energetically less favored, and that the CRI
model faces very severe problems of energetics (Ramaty et al.
1998b).

\begin{figure}

\psfig{file=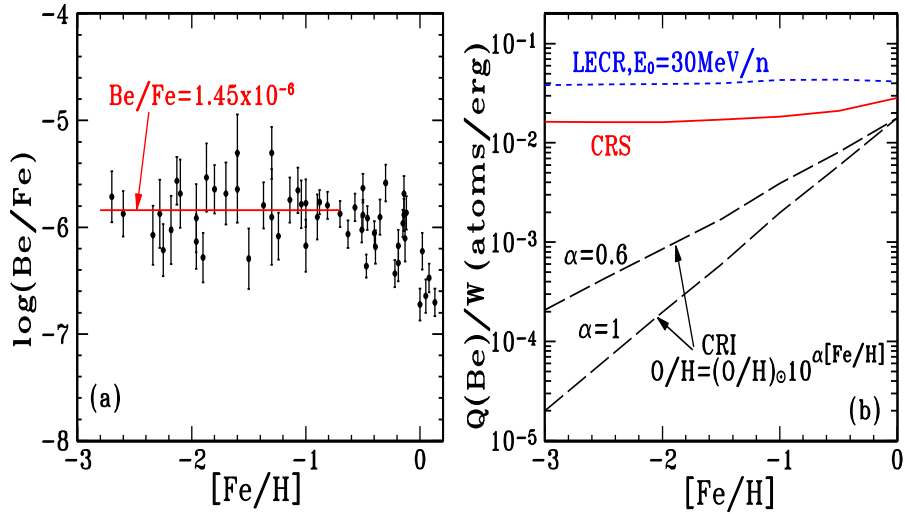,height=7cm,width=12cm} \caption[]{Panel (a):
observed Be-Fe abundance ratio as a function of [Fe/H]; data
compilation by Vangioni-Flam et al. (1998). Panel (b): number of
Be atoms produced per erg of cosmic-ray source kinetic energy; the
ambient medium is neutral and the cosmic-ray escape
length from the Galaxy is 10 g cm$^{-2}$. If no cosmic-ray escape
is allowed (closed Galaxy), Q(Be)/W would increase by a
factor of $\simeq$2, except for the LECR model for which most of the
accelerated particles stop in 10 g cm$^{-2}$. The CRI curve for
$\alpha = 0.6$ takes into account the less rapid decrease of O/H at
low [Fe/H] (Israelian et al. 1998).}
\end{figure}

The three models (CRI, CRS, LECR) imply different current epoch
cosmic-ray origin scenarios. While the CRI scenario posits
acceleration from the ambient ISM (Ellison, Drury \& Meyer 1997),
the CRS model implies that the cosmic rays are accelerated out of
fresh supernova ejecta. We have shown (Lingenfelter et al. 1998)
that the standard arguments against such a cosmic-ray origin (Webber
1997; Meyer et al. 1998) can be answered, and that the most likely
scenario involves the collective acceleration by successive
supernova shocks of ejecta-enriched matter in the interiors of
superbubbles (Higdon et al. 1998). This scenario is based on
observations (summarized by Higdon et al. 1998) showing that most of
the Type II and Ibc supernova progenitors (O and B stars) are
produced in giant OB associations, that the subsequent supernova
explosions produce giant superbubbles that make up the hot, low
density phase filling roughly half of the ISM, and that the bulk (80
to 90\%) of all supernovae occur in these superbubbles enabling
their shocks to mostly accelerate fresh ejecta matter. These
results, by themselves and quite apart from the  LiBeB origin
arguments, favor the CRS over the CRI scenario for cosmic ray
origin. Independent arguments that the cosmic rays are accelerated
from supernova ejecta were given by Erlykin \& Wolfendale (1997).

The LECR scenario was motivated by the reported (Bloemen et al.
1994) detection with COMPTEL/CGRO of C and O nuclear gamma-ray lines
from Orion. These gamma rays were attributed to a low energy
cosmic-ray component highly enriched in C and O relative to protons
and $\alpha$ particles (see Ramaty 1996 for review and Ramaty et al.
1996 for extensive calculations of LiBeB production by LECRs). It
was suggested (Bykov 1995; Ramaty et al. 1996; Parizot, Cass\'e, \&
Vangioni-Flam 1997) that such enriched LECRs might be accelerated
out of metal-rich winds of massive stars and the ejecta of
supernovae from massive star progenitors which explode within the
bubble around the star formation region due to their very short
lifetimes. But since the validity of these Orion observations has
been questioned by the COMPTEL team (private communication, V.
Sch\"onfelder, 1998), the determination of the role of LECRs in
LiBeB origin must await future nuclear gamma-ray line observations.

\section{The Energetics of Be Production}

We calculate the energy in cosmic rays, $W$, needed to produce a
given number of Be atoms, $Q({\rm Be})$ (Ramaty et al. 1997). The
calculation assumes a cosmic-ray source generating accelerated
particles with given composition and energy spectrum, which then
propagate and interact in an ambient medium of given composition.
The transport of the particles is characterized by a target
thickness, $X$, measured in g cm$^{-2}$. Results are shown in
Figure~1b for a neutral ambient medium and $X=10~{\rm g~cm}^{-2}$,
the approximate Galactic target thickness for the current epoch
cosmic rays. The accelerated particle source energy spectra are
taken proportional to $(p^{-2.2}/\beta) e^{-E/E_0}$, where $p$,
$c\beta$ and $E$ are particle momentum, velocity and
energy/nucleon, respectively; except for the LECR case, $E_0$ is
ultrarelativistic. For both the CRS and LECR cases, the
accelerated particle composition is independent of [Fe/H] and the
same as that of the current epoch cosmic rays, except that there
are no protons and $\alpha$-particles for the latter. The ambient
medium composition is solar scaled with Fe/H, except that for the
CRI case we also consider a slower decrease of the O abundance,
O/H=(O/H)$_\odot$$10^{0.6{\rm [Fe/H]}}$, which fits the recent
Israelian et al. (1998) data. Such a modification of the ambient
medium abundances has only a negligible effect on the calculations
for the CRS and LECR models. The accelerated particle composition
for the CRI case varies with [Fe/H], being equal to the ambient
medium abundances increased by factors consistent with ISM shock
acceleration theory (Ellison et al. 1997). For both the CRS and
LECR cases $Q({\rm Be})/W$ is essentially constant. On the other
hand, for the CRI case $Q({\rm Be})/W$ decreases with decreasing
[Fe/H], becoming very low at low [Fe/H] in spite of the increase
by as much as an order of magnitude due to the incorporation of
the enhanced ISM O abundance.

The required energy per SNII is given by
\begin{eqnarray}
W_{\rm SN} = {Q_{\rm SN}({\rm Be})/(Q({\rm Be})/W)}~~,
\end{eqnarray}
independent of the details of the employed Galactic chemical
evolution model. Thus, for any given cosmic-ray scenario, the main
uncertainty is due to $Q_{\rm SN}({\rm Be})$, the Be yield per
SNII. As Figure~1a indicates that the Be and Fe yields should be
well correlated, we take $Q_{\rm SN}({\rm Be})=({\rm Be}/{\rm
Fe})Q_{\rm SN}({\rm Fe})$, where $Q_{\rm SN}({\rm Fe})$ is the
number of Fe nuclei ejected per SNII. The problem then is the
determination of this number.

\begin{figure}
\hskip 1cm \psfig{file=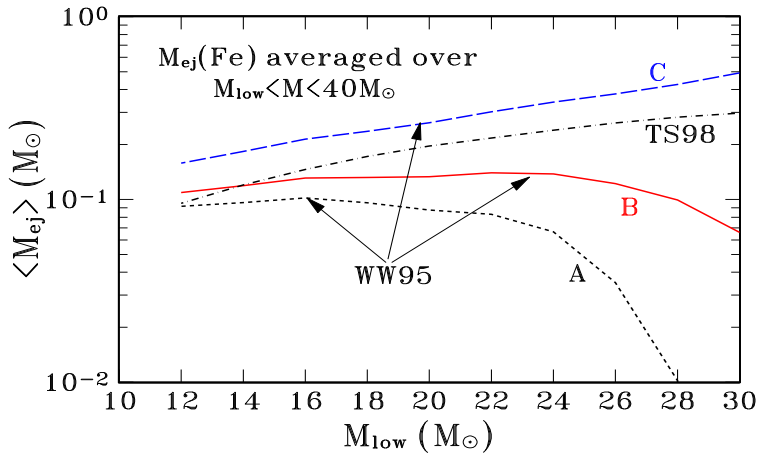,height=6cm,width=10cm}
\caption[]{
IMF averaged Fe mass ejected mostly as $^{56}$Ni per SNII for
progenitor masses in the range M$_{\rm low}$ to 40 M$_\odot$. Curves
A, B and C, corresponding to different final ejecta kinetic
energies, are based on the calculations of Woosley \& Weaver (1995)
for metallicity 10$^{-4}$. The TS98 curve employs the results of
Tsujimoto \& Shigeyama (1998). }
\end{figure}

Using the calculations of Woosley \& Weaver (1995, WW95), we
calculate the ejected Fe mass (mostly from $^{56}$Ni) per SNII
averaged over the Salpeter IMF for progenitor masses M$_{\rm
low}$$<$M$<$40 M$_\odot$. The results are shown by curves A, B and C
in Figure~2. These correspond to the WW95 cases A, B and C which
give different $^{56}$Ni yields for progenitor masses above 30
M$_\odot$, due to different assumed final ejecta kinetic energies,
typically 1.2, 2 and 2.5$\times$10$^{51}$ ergs for cases A, B and C,
respectively. Also shown in Figure~2 (the TS98 curve) is a similar
average based on the results of Tsujimoto \& Shigeyama (1998). We
see that the ejected mass averaged over the entire 10 to 40
M$_\odot$ range is about 0.1 M$_\odot$ for all four cases. Taking
into account the main sequence lifetimes of the SNII progenitors in
this mass range, such an average would be appropriate for
evolutionary scenarios in which [Fe/H] reaches 10$^{-3}$ in 10 Myrs
or more. Since this is quite reasonable (e.g. Ramaty et al. 1998b),
we shall use 0.1 M$_\odot$ in our subsequent estimates. However we
note that if [Fe/H]=10$^{-3}$ is reached in just a few Myrs, only
SNIIs from progenitors more massive than about 25 M$_\odot$ can
contribute, allowing ejected Fe masses lower than 0.1 M$_\odot$, but
only for case A.

Combining the average ejected Fe mass of 0.1 M$_\odot$ with the
constant Be/Fe (Figure~1), we obtain the required Be yield per SNII,
$3\times 10^{48}$ atoms. As the recent analysis of Fields \& Olive
(1998) indicates somewhat lower Be/Fe values at the lowest [Fe/H],
we assign a downward uncertainty to this value of about a factor of
3. Using $Q_{\rm SN}({\rm Be})\simeq 3\times 10^{48}$ in Equation
(1), we obtain $W_{\rm SN}({\rm CRS}) \simeq 1.5\times 10^{50}$ erg,
which as already mentioned is in excellent agreement with the
current epoch value. On the other hand, using the $\alpha$=0.6 curve
in Figure~1b at [Fe/H]$ = 10^{-3}$, we obtain $W_{\rm SN}({\rm CRI})
\simeq 1.5\times 10^{52}$ erg, a highly excessive value, even if it
were possible to reduce it by the above mentioned factor of 3. For
the LECR model of Vangioni-Flam et al. (1996), in which only the
$>$60 M$_\odot$ progenitors contribute, we obtain $W_{\rm SN}({\rm
LECR}) \simeq 8\times 10^{50}$ erg. This energy, however, is just
that residing in the metals. If protons and $\alpha$ particles
accompany the metals with abundances equal to those of the current
epoch cosmic rays, $W_{\rm SN}({\rm LECR}) \simeq 5\times 10^{51}$
erg for this model. But this energetic efficiency  can be improved
by relaxing the $>$60 M$_\odot$ progenitor constraint, and by
allowing $E_0$ (defined above) to exceed 30 MeV/nucleon but still be
nonrelativistic. Observations of Galaxy-wide nuclear gamma-ray lines
are needed to determine the contribution of the LECR component to
LiBeB production.

RR wishes to acknowledge Sean Scully for useful discussions.

\end{document}